\documentclass[10pt,conference]{IEEEtran}
\IEEEoverridecommandlockouts
\usepackage{multirow}
\usepackage[table,xcdraw]{xcolor}
\usepackage[bottom]{footmisc}
\usepackage{float}
\usepackage{cite}
\usepackage{bm}
\usepackage{amsmath,amssymb,amsfonts}
\usepackage{graphicx}
\usepackage{textcomp}
\usepackage[caption=false,font=footnotesize]{subfig}
\usepackage[most]{tcolorbox}
\usepackage{xcolor}
\usepackage{multirow}
\usepackage{booktabs}
\usepackage{amsfonts}
\usepackage{amsmath}
\usepackage{siunitx}
\usepackage{mwe}
\usepackage[inline]{enumitem}
\usepackage[utf8]{inputenc}
\usepackage[english]{babel}
\usepackage[linesnumbered,ruled,vlined]{algorithm2e}
\SetCommentSty{mycommfont}
\SetKwInput{KwInput}{Input}                
\SetKwInput{KwOutput}{Output}              
\usepackage[noend]{algpseudocode}
\usepackage{balance}
\usepackage{url}
\setlist[enumerate]{nosep}
\usepackage[normalem]{ulem}
\usepackage{xspace}
\newcommand{\etal}{et al.\xspace}
\newcommand{\DRLM}{\texttt{DRLM}\xspace}
\usepackage{caption}

\begin{document}
\title{DRLM: Deep Reinforcement Learning-Based\\LLM Query Orchestration in Edge Environments\vspace*{-3mm}}
\author{
\IEEEauthorblockN{
Reza Farahani\IEEEauthorrefmark{1},
Zoha Azimi Ourimi\IEEEauthorrefmark{2},
Mario Colosi\IEEEauthorrefmark{3},
Lauri Lov\'en\IEEEauthorrefmark{4},
Christian Timmerer\IEEEauthorrefmark{2},
Schahram Dustdar\IEEEauthorrefmark{1}
}
\IEEEauthorblockA{\IEEEauthorrefmark{1}
Distributed Systems Group (DSG), TU Wien, Austria}
\IEEEauthorblockA{\IEEEauthorrefmark{2}
Christian Doppler Laboratory ATHENA, Department of Information Technology (ITEC), University of Klagenfurt, Austria}
\IEEEauthorblockA{\IEEEauthorrefmark{3}
MIFT Department, University of Messina, Italy}
\IEEEauthorblockA{\IEEEauthorrefmark{4}
Center for Ubiquitous Computing, University of Oulu, Finland}}
\maketitle
\begin{abstract}
Large language model (LLM) services increasingly process heterogeneous queries with diverse latency, accuracy, and resource requirements. While edge deployment reduces response time, the heterogeneity of devices and the diversity of model families, parameter scales, and quantization levels make efficient LLM query orchestration challenging. This paper introduces \DRLM, a \underline{D}eep \underline{R}einforcement Learning-based L\underline{L}\underline{M} query orchestration framework in edge environments. \DRLM integrates two lightweight predictors: \emph{(i)} a class-conditioned quality estimator that maps queries to semantic categories and infers model performance, and \emph{(ii)} a feature-driven latency predictor that estimates inference time across model-device configurations. These predictions, combined with system state, feed a factorized Proximal Policy Optimization (PPO) agent that performs state-aware orchestration decisions. To enable data-driven orchestration, we construct a large-scale benchmarking dataset with \num{223835} measurements spanning \num{1258} queries, \num{6} query classes, \num{8} model families (\num{32} deployed instances), \num{5} quantization levels, and heterogeneous edge devices.  Evaluation on a \num{64}-node edge cluster and comparison with three baselines and two state-of-the-art methods show that \DRLM reduces inference latency by up to \qty{51}{\percent} and queuing delay by up to \qty{67}{\percent}, while incurring at most \qty{8}{\percent} accuracy loss. It improves latency under increasing workloads up to \qty{61.4}{\percent}, demonstrating robust and stable orchestration.
\end{abstract}

\vspace{-.2cm}
\section{Introduction}
\label{sec:Introduction }
Large language models (LLMs) power modern AI services like chatbots and code-generation systems, processing queries with diverse structures, reasoning at varying levels of complexity, and meeting varying latency requirements~\cite{farahani2026towards}. In edge environments, this challenge is amplified by hardware heterogeneity, where model instances, spanning different families, parameters, and quantizations, show distinct latency-quality-resource trade-offs~\cite{farahani2026clusterless, azimi2026ellmpeg,loven2025agentic}. Thus, LLM orchestration becomes a fine-grained, per-query decision problem over model-device-quantization configurations rather than a simple model or device selection task~\cite{farahani2026lmedge,farahaniserverless}.
To quantify these interactions, we evaluate representative configurations on two edge devices: a constrained Raspberry Pi 4 (RP) and a more capable NVIDIA Jetson ORIN AGX (JOA). Each configuration is denoted as $(x,y)$, where $x$ is the model size (in billions of parameters) and $y$ is the quantization level. Fig.~\ref{fig:motiv} reports average accuracy and response time for \texttt{Mathematics}~\cite{cobbe2021gsm8k} and \texttt{TruthfulQA}~\cite{truthful} queries, revealing three key observations:
%%%
\begin{figure}[!t]
    \centering
    \subfloat[Average accuracy on RP.\label{SLM-acc}]{
        \includegraphics[width=0.48\linewidth]{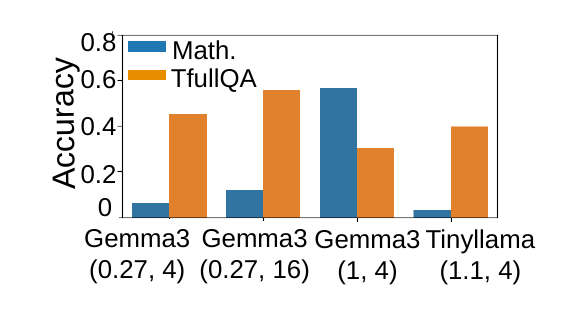}}
    \hfill
    \subfloat[Average accuracy on JOA.\label{LLM-acc}]{
        \includegraphics[width=0.48\linewidth]{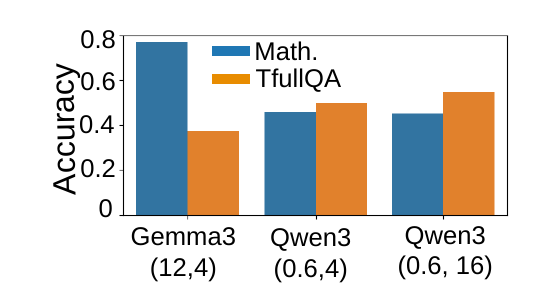}}
    \vspace{-2ex}
    \subfloat[Average response time on RP.\label{SLM-time}]{
        \includegraphics[width=0.48\linewidth]{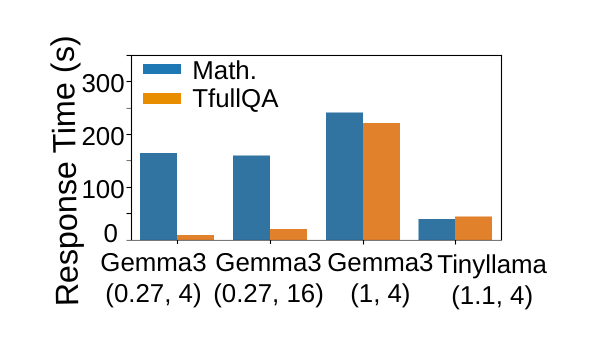}}
    \hfill
    \subfloat[Average response time on JOA.\label{LLM-time}]{
        \includegraphics[width=0.48\linewidth]{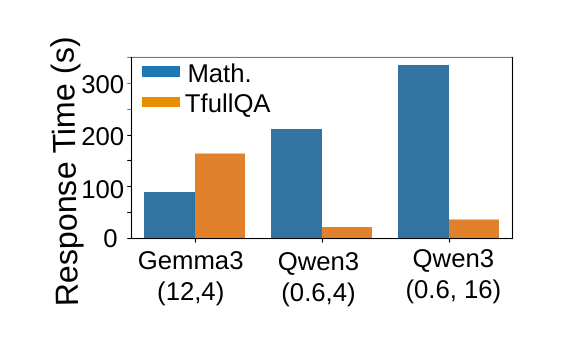}}
        \vspace{-5pt}
    \caption{\small{Latency-quality of LLM configurations on \texttt{RP} and \texttt{JOA} edge devices for \texttt{Mathematics} and \texttt{TruthfulQA} queries.}}
    \label{fig:motiv}
\end{figure}
\subsubsection{Model-family impact}
% Across fixed devices and comparable configurations, model families exhibit distinct, query-dependent behavior. 
On JOA (Fig.~\ref{LLM-acc}), \texttt{Gemma3}$(12,4)$ has the highest accuracy on \texttt{Mathematics}, whereas \texttt{Qwen3}$(0.6,4)$ and $(0.6,16)$ outperform others on \texttt{TruthfulQA}. This indicates that model architecture, not just size, determines task-specific performance.
\subsubsection{Scaling impact}
On RP (Fig.~\ref{SLM-acc}), \texttt{Gemma3}$(1,4)$ improves \texttt{Mathematics} accuracy over $(0.27,4)$ but degrades performance on \texttt{TruthfulQA}. Similarly, on JOA, \texttt{Gemma3}$(12,4)$ improves \texttt{Mathematics} but does not consistently dominate across tasks. Thus, scaling benefits are workload-dependent rather than monotonic.
\subsubsection{Quantization impact}
On RP (Figs.~\ref{SLM-acc}-\ref{SLM-time}), \texttt{Gemma3}$(0.27,16)$ improves \texttt{TruthfulQA} accuracy over $(0.27,4)$ but increases latency. On JOA (Fig.~\ref{LLM-time}), \texttt{Qwen3}$(0.6,16)$ incurs higher latency than $(0.6,4)$ with marginal accuracy gains. 
These results show that quantization effects are highly configuration- and device-dependent, and must be selected jointly with model and system characteristics rather than assumed to improve latency or accuracy.
\begin{figure}[H]
\vspace{-12pt}
\centering
\begin{tcolorbox}[colback=gray!10,colframe=black!40!black,
        boxrule=0.8pt,arc=3pt,left=6pt,right=6pt,top=3pt,bottom=3pt]
\textbf{Implication:} Effective LLM orchestration on edge clusters needs \emph{joint, state-aware modeling} of query semantics, model configurations, and runtime dynamics, which cannot be captured by static or single-objective policies.
\end{tcolorbox}
\vspace{-15pt}
\end{figure}

Existing approaches address LLM query orchestration from largely decoupled perspectives. Inference-aware systems such as ExeGPT~\cite{oh2024exegpt} optimize execution by modeling input/output token lengths to control batching and parallelism, while recent schedulers such as Bullet~\cite{lin2026bullet} improve throughput and latency through KV-cache-aware memory management and dynamic prefill-decode co-scheduling. At the infrastructure level, DynamoLLM~\cite{stojkovic2025dynamollm} adapts system parameters such as instance scaling, parallelism, and GPU frequency to meet latency constraints under dynamic workloads. Model selection has been explored via learning and optimization. 
OptLLM~\cite{liu2024optllm} selects models from a Pareto frontier based on predicted performance under fixed constraints. 
RouteLLM~\cite{ong2025routellm} reduces routing to a binary decision between a fixed strong-weak model pair. Recent extensions incorporate system-level realism: RouterWise~\cite{kasnavieh2026routerwise} jointly models query routing and resource allocation, Robust Batch-Level Routing~\cite{markovic2026robust} enforces cost and capacity constraints at the batch level, and TRouter~\cite{liu2026trouter} improves robustness under cold-start query distributions. Edge-oriented works focus on deployment and resource distribution rather than per-query orchestration. Jang~\etal~\cite{jang2025edge} proposes device-level selection strategies to balance load, while Edge-LLM~\cite{cai2024edge} and EdgeShard~\cite{zhang2024edgeshard} optimize collaborative inference and model partitioning across devices. 
\begin{figure}[H]
\vspace{-10pt}
\centering
\begin{tcolorbox}[colback=gray!10,colframe=black!40!black,
        boxrule=0.8pt,arc=3pt,left=6pt,right=6pt,top=4pt,bottom=4pt]
\textbf{State-of-the-art limitations:} Existing solutions either optimize execution for fixed models or perform simplified model selection, treating latency and quality as static. They do not jointly capture \emph{query semantics}, \emph{model configuration} (family, scale, quantization), and \emph{dynamic system state} (utilization, queueing), leading to load-coupled, query-dependent behavior and limiting fine-grained per-query orchestration in heterogeneous edge environments.
\end{tcolorbox}
\vspace{-15pt}
\end{figure}
%%%%%%%%

This paper introduces \DRLM, a \underline{D}eep \underline{R}einforcement Learning-based framework for fine-grained L\underline{L}\underline{M} query orchestration in heterogeneous edge clusters. \DRLM leverages \emph{(i)} a query profiler that extracts semantic and structural features, \emph{(ii)} lightweight predictors that estimate class-conditioned quality and configuration-dependent latency, and \emph{(iii)} a factorized Proximal
Policy Optimization (PPO)-based agent that combines these predictions with runtime system state information (e.g., resource utilization and queue dynamics) to make per-query orchestration decisions. To enable data-driven learning, we construct a large-scale benchmarking dataset with \num{223835} measurements over \num{1258} queries and \num{6} classes, capturing fine-grained query-model-device interactions. 
We evaluate \DRLM on a realistic Kubernetes-based edge cluster with \num{64} heterogeneous nodes and \num{32} LLM instances spanning \num{8} model families, \num{11} parameter scales, and \num{5} quantization levels. Comparison with three baselines and two state-of-the-art methods shows that \DRLM reduces inference latency by up to \qty{51}{\percent} and queuing delay by up to \qty{67}{\percent}, while incurring at most \qty{8}{\percent} accuracy loss. Under increasing workload, it further reduces latency by up to \qty{61.4}{\percent}, showing robust orchestration.
% The paper is organized as follows. Section~\ref{sec:Formulation} formulates the LLM query orchestration problem as a Markov decision process. Section~\ref{sec:design} designs the modular system architecture of \DRLM. Section~\ref{sec:EvaluationSetup} describes the experimental setup, including testbed, datasets, methods, and evaluation metrics, followed by the results in Section~\ref{sec:EvaluationSetup}. Section~\ref{sec:Conclusion} concludes the paper and outlines future work.

\section{Problem Formulation}
\label{sec:Formulation}
We formulate LLM query orchestration in edge clusters as a \emph{Markov Decision Process} (MDP) defined by the tuple \mbox{$(\mathcal{S}, \mathcal{A}, \pi, r)$}, where $\mathcal{S}$ is the state space, $\mathcal{A}$ the action space, $\pi$ the policy, and $r$ the reward function. At each decision step, the agent observes the system state and assigns an incoming query to a model-device configuration.
\subsubsection{Environment}
We consider an edge cluster with $N$ heterogeneous devices \mbox{$\mathcal{E}=\{e_1,\dots,e_N\}$}, hosting $K$ deployed LLM instances \mbox{$\mathcal{M}=\{m_1,\dots,m_K\}$}. Each instance \mbox{$m_i\in\mathcal{M}$} corresponds to a unique configuration \mbox{$(f,p,z)$}, where \mbox{$f\in\mathcal{F}$} denotes the model family, \mbox{$p\in\mathcal{P}$} the parameter scale, and \mbox{$z\in\mathcal{Z}$} the quantization level. Thus, configurations differing in family, size, or quantization are treated as distinct deployable instances. Incoming queries \mbox{$\mathcal{X}=\{x_1,x_2,\dots\}$} arrive online according to a stochastic arrival process. Each query $x_i$ must be assigned to a feasible model-device pair $(m,e)$, subject to deployment and resource constraints.
To enable efficient decision-making, the environment provides predictive feedback via lightweight models: \emph{(i)} inference latency \mbox{$\widehat{t}_{inf} = \mathcal{T}(x_i,m,e)$} and response quality \mbox{$\widehat{y} = \mathcal{A}(x_i,m)$}. These predictors approximate execution outcomes without performing full inference, enabling fast candidate evaluation.
\subsubsection{State space}
We define the state \mbox{$s_i\in\mathcal{S}$} at the arrival of query $x_i$ as
\mbox{$s_i=\big[ \phi(x_i),\; \widehat{\mathbf{y}},\; \widehat{\mathbf{t}}_{inf},\; \mathbf{u},\; \mathbf{q} \big]$}, where $\phi(x_i)$ denotes the extracted query representation capturing semantic and structural properties. 
The vectors \mbox{$\mathbf{u}=\{u(e)\}_{e\in\mathcal{E}}$} and \mbox{$\mathbf{q}=\{q(e)\}_{e\in\mathcal{E}}$} capture device resource utilization and queue state, respectively, where $q(e)$ encodes the queue length and remaining processing time of active requests. This state representation enables joint reasoning over query characteristics, model behavior, and system dynamics.
\begin{figure*}[!t]
    \centering
    \includegraphics[width=.9\linewidth]{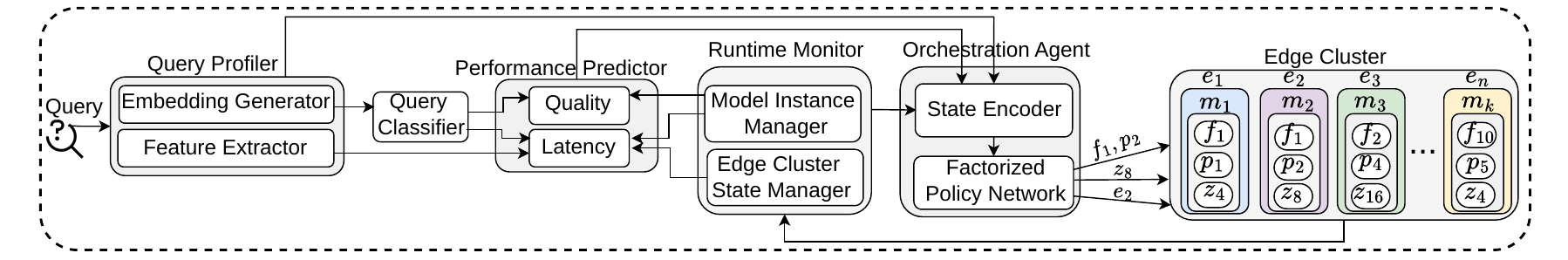}
    \vspace{-.3cm}
    \caption{\small{\DRLM architecture; models are indicated by family $f$, parameter scale $p$, and quantization level $z$, and executed on edge device $e$.}}
    \vspace{-.4cm}
    \label{arch}
\end{figure*}
\subsubsection{Action space}
The action space $\mathcal{A}$ comprises feasible assignments \mbox{$a=(m,e)$}, selecting a model configuration and execution device, constrained by deployment placement and device compatibility.
\subsubsection{Reward function} The reward balances response quality and end-to-end serving latency via
$r(s_i,a) = \alpha \cdot \widehat{\bar{y}} - \beta \cdot \Phi(\widehat{t}_{e2e}),$ where $\widehat{\bar{y}}$ is the normalized predicted response quality and $\widehat{t}_{e2e}$ is the end-to-end latency defined as
$\widehat{t}_{e2e} = \widehat{t}_{inf}(x_i,m,e) + t_{wait}(e),$ where $t_{wait}(e)$ denotes the waiting time before query $x_i$ can start execution on device $e$, as induced by the current queue state and ongoing executions. The function $\Phi$ is a normalized exponential penalty that increases the cost of high-latency assignments more aggressively than a linear formulation. Both $\widehat{\bar{y}}$ and $\widehat{t}_{e2e}$ are normalized via min-max scaling across candidate actions. The coefficients $\alpha$ and $\beta$ control the trade-off between response quality and latency.
\subsubsection{Policy learning}
We learn a stochastic policy $\pi_\theta(a|s)$ using PPO, which maximizes the expected cumulative reward. At each step, the agent samples an action \mbox{$a_i=(m,e)$} from \mbox{$\pi_\theta(a|s_i)$} and receives reward \mbox{$r(s_i,a_i)$}. 
The policy is optimized using the PPO objective:
\begin{equation}
\begin{aligned}
\mathcal{L}(\theta) = \mathbb{E}\Big[
\min\big(
\rho_i(\theta)\hat{A}_i,\;
\mathrm{clip}(\rho_i(\theta), 1-\epsilon, 1+\epsilon)\hat{A}_i
\big)
\Big] \\
- c_1 \mathcal{L}_{\text{value}}
+ c_2 \mathcal{H}(\pi_\theta)
\end{aligned}
\end{equation}
where \mbox{$\rho_i(\theta)=\frac{\pi_\theta(a_i|s_i)}{\pi_{\theta_{\text{old}}}(a_i|s_i)}$, $\hat{A}_i$} is the advantage estimate, \mbox{$\mathcal{L}_{\text{value}}$} is the value-function loss, and \mbox{$\mathcal{H}(\pi_\theta)$} denotes the entropy bonus encouraging exploration. The coefficients $c_1$ and $c_2$ control the contribution of the value loss and entropy regularization. We employ an actor-critic architecture, where the policy \mbox{$\pi_\theta$} (actor) is trained jointly with a value function $V_\psi(s)$ (critic). Training relies on trajectories generated by the predictors $\mathcal{T}$ and $\mathcal{A}$, avoiding execution of real LLM inference and enabling efficient learning under dynamic system conditions.

\section{\DRLM Design}
\label{sec:design}
Fig.~\ref{arch} illustrates the \DRLM architecture for fine-grained LLM query orchestration in heterogeneous edge clusters. The system consists of five modules that jointly enable efficient per-query model-device selection.
%%%%%%%%%%
\subsubsection{Query profiler} transforms each incoming query into a structured representation capturing both structural and semantic properties. This includes lightweight features (e.g., token length, syntactic complexity) and a dense embedding generated via a compact sentence encoder.
%%%%%%%%%%
\subsubsection{Query classifier} maps each incoming query to a semantic class distribution over a set of query categories. Implemented using a lightweight encoder-classifier pipeline, it provides a low-dimensional abstraction of query intent. For low-confidence or out-of-distribution inputs, the system falls back to a default class associated with robust model configurations, ensuring stable behavior under uncertainty. The class set can be extended by retraining on new workloads. This abstraction reduces prediction variance while preserving query-dependent performance patterns.
%%%%%%%%%%
\subsubsection{Performance predictor} estimates query-dependent model behavior, enabling fast evaluation of candidate model-device assignments without executing inference.
\paragraph{Quality} infers expected model performance from query classes, avoiding noisy per-query accuracy prediction. It captures consistent patterns across semantic categories, enabling stable quality estimation.
\paragraph{Latency} predicts inference time using
query features, model configuration, and device characteristics. It captures execution behavior across devices and configurations, enabling inference-time estimation under varying system conditions.
%%%%%%%%%%
\subsubsection{Runtime monitor}
The runtime monitor maintains the dynamic system state required for orchestration.
\paragraph{Model instance manager}
tracks the deployment and availability of model instances across devices.
\paragraph{Edge cluster state manager}
continuously collects device-level metrics, including resource utilization and execution state. In particular, it tracks current load and remaining processing time of active requests, forming the system state $(\mathbf{u}, \mathbf{q})$ used by the orchestration agent.
%%%%%%%%%%
\subsubsection{Orchestration agent}
The agent integrates predictive signals and system state to perform query-conditioned decisions.
\paragraph{State encoder}
Constructs the decision state by combining query representation, predicted performance, and runtime state into a compact latent representation.
%%
%%%%%%%%%%%%%
\begin{figure*}[!t]
    \centering
    \subfloat[Prompt complexity (KDE).\label{dataset-complexity}]{
        \includegraphics[width=0.24\linewidth]{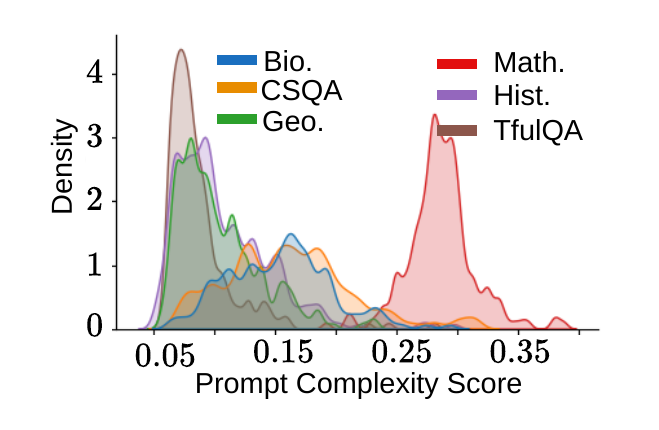}}
    \hfill
    \subfloat[Prompt token count (KDE).\label{dataset-token}]{
        \includegraphics[width=0.24\linewidth]{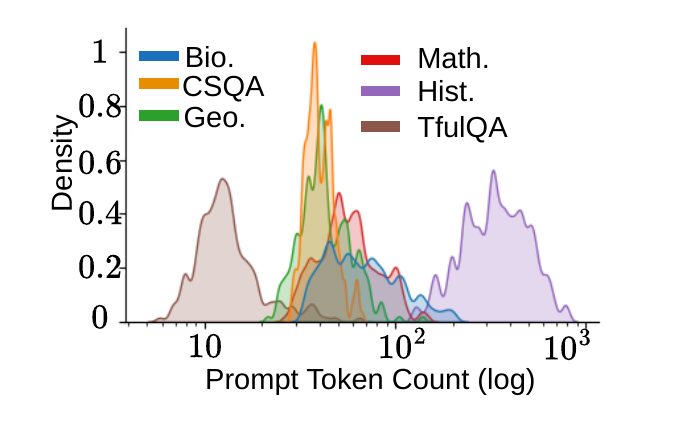}}
    \subfloat[Accuracy.\label{dataset-accuracy}]{
        \includegraphics[width=0.24\linewidth]{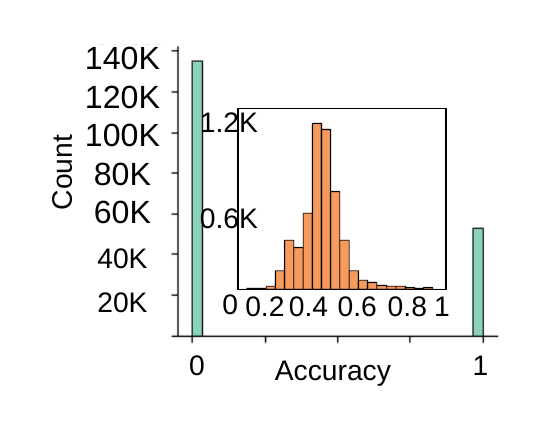}}
    \hfill
    \subfloat[Inference time.\label{dataset-inf}]{
        \includegraphics[width=0.24\linewidth]{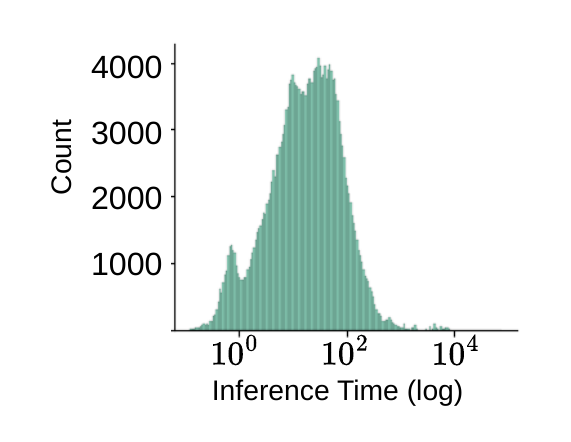}}
        \vspace{-5pt}
    \caption{\small{Distribution of query, accuracy, inference time in the benchmarking dataset.}}
    \vspace{-.6cm}
    \label{fig:benchmark}
\end{figure*}
%%%%%%%%%%%%%%%%
\paragraph{Factorized policy network} learns a stochastic mapping \mbox{$\pi_\theta(a|s)$} over model-device assignments. To handle the combinatorial action space, we adopt an autoregressive factorization method over model configuration and device selection:
\begin{equation}
\begin{aligned}
\pi_\theta(a|s_i)=\pi_f(f,p|s_i)\cdot \pi_z(z|s_i,f,p)\cdot \pi_e(e|s_i,f,p,z),
\end{aligned}
\end{equation}
where $(f,p,z)$ defines the selected model instance $m$ with family $f$, parameter scale $p$ and quantization level $z$. The policy is implemented using separate output heads for each decision stage (model family/scale, quantization, and device), each conditioned on the state and preceding selections. Invalid choices are dynamically masked based on deployment and compatibility constraints. This factorization reduces the effective action space and improves sample efficiency by enabling structured credit assignment across model-device selection.
%%%%%%%%%%
% \subsubsection{Execution flow} Upon query arrival, the \emph{profiler} extracts structural and semantic representations. The \emph{classifier} maps the query to a semantic distribution. The \emph{performance predictor} then evaluates candidate model-device pairs using class-conditioned quality and feature-based latency estimation. The \emph{runtime monitor} provides the current system state, including resource utilization and execution status. These inputs are then aggregated by the \emph{orchestration agent}, which selects the target model-device pair for execution.
%%%%%%%%%%%%

\section{Evaluation Setup}
\label{sec:EvaluationSetup}
This section describes the setup, including the testbed, model deployment, workload, and evaluation methodology.
\subsubsection{Edge testbed}
We evaluate \DRLM on an edge cluster comprising \num{64} heterogeneous nodes across \num{5} device classes: \texttt{Raspberry Pi 3/4} (RP, \numrange{2}{12} CPU cores, \qtyrange{1}{4}{GB} RAM, \num{38} nodes), \texttt{Jetson Nano} (NJN, \num{4} cores, \qty{4}{GB} RAM, \num{128}-core GPU, \num{6} nodes), \texttt{Jetson Orin Nano} (JON, \num{6} cores, \qty{8}{GB} RAM, \num{1024}-core GPU, \num{2} nodes), \texttt{Jetson Orin AGX} (JOA, \num{12} cores, \qty{64}{GB} RAM, \num{2048}-core GPU, \num{2} nodes), and \texttt{KVM}-based VMs (\numrange{2}{24} vCPU, \qtyrange{16}{32}{GB} RAM, \num{16} nodes). The cluster runs \texttt{Kubernetes} with \texttt{containerd}, hosting containerized LLM inferences.
 We collect runtime telemetry using \texttt{Prometheus} and \texttt{cAdvisor} for CPU/memory, and \texttt{tegrastats} for GPU utilization on Jetson devices. Metrics are aggregated to construct system state, including utilization, active load, and processing backlog. All model instances are exposed via REST, enabling dynamic query dispatch to valid model-device pairs.
\begin{table}[!t]
\centering
\fontsize{6.5pt}{6.5pt}\selectfont
\caption{\small{Summary of deployed models on the edge devices.}}
\label{tab:llm-depl}
\vspace{-5pt}
\renewcommand{\arraystretch}{1.3}
\begin{tabular}{|c|c|c|c|}
\hline
\emph{Model family} & \emph{Parameter scale} & \emph{Quantization}                    & \emph{Edge device}                 \\ \hline
GPT-OSS      & 20 B       & $z_4$                           & JOA, VM                \\ \hline
Gemma2       & 2B         & $z_4$                           & NJN, RP, JOA, JON, VM \\ \hline
Gemma3       & 0.27B      & $z_4, z_{16}$         & NJN, RP, JOA, JON, VM \\ \hline
Gemma3       & 12B        & $z_4$                        & JOA, VM                \\ \hline
Gemma3       & 1B         & $z_3,z_4$    & NJN, RP, JOA, JON, VM \\ \hline
Llama3.1     & 8B         & $z_8, z_{16}$                      & JOA, VM                \\ \hline
Llama3.2     & 1B         & $z_4, z_8$                 & NJN, RP, JOA, JON, VM \\ \hline
Llama3.2  & 3B   & $z_2, z_3, z_4$    & RP, JOA, JON, VM      \\ \hline
Qwen3        & 0.6B       & $z_4, z_8, z_{16}$ & JOA, VM                \\ \hline
Qwen3        & 1.7B       & $z_4$                        & JOA, VM                \\ \hline
Tinyllama & 1.1B & $z_2, z_3, z_4, z_8$ & NJN, RP, JOA, JON, VM \\ \hline
Mistral      & 7B         & $z_8$                           & JOA, VM                \\ \hline
\end{tabular}
\end{table}
%%%
\subsubsection{LLM models}
We deploy \num{32} LLM instances from \num{8} model families, spanning parameters from \qty{0.27}{B} to \qty{20}{B} with diverse quantizations, resulting in a heterogeneous space of model-configuration variants. Our deployment is device-aware, i.e., highly quantized lightweight models are placed on constrained devices (RP, NJN), while larger and higher-precision models are hosted on more capable nodes (JON, JOA, VM). This induces diverse latency-quality trade-offs and non-uniform actions. Table~\ref{tab:llm-depl} summarizes the deployment.
%%%%%%%%%%%%
% \begin{table}[!t]
%     \centering    
%     \fontsize{6.5pt}{6.5pt}\selectfont
%     \caption{\small{Summary of the \num{1258} in \num{6} categories.}}
%     \label{tab:dataset}
%     \vspace{-5pt}
%     \begin{tabular}{|l|l|c|l|}
% \hline
% \textit{Dataset} &  \textit{Category} &  \textit{Size} & \textit{Response} \\ \hline
% \multirow{3}{*}{MMLU~\cite{hendrycks2020measuring}}  
% & College biology (Bio) & \num{165} & Multi-choice \\ \cline{2-4}
% & World history (Hist)  & \num{268} & Multi-choice \\ \cline{2-4}
% & Geography (Geo)      & \num{225} & Multi-choice \\ \hline
% GSM8K~\cite{cobbe2021gsm8k} & Math problems (Math) & \num{200} & Explanation \\ \hline
% CommonsenseQA~\cite{talmor2018commonsenseqa} & Commonsense reasoning (CSQA) & \num{200} & Multi-choice \\ \hline
% TruthfulQA~\cite{truthful} & Misconception QA (TfulQA)& 200 & Explanation \\ \hline
% \end{tabular}
% \end{table}
%%%%%%%%%%%%%%
\subsubsection{Query workload}
We evaluate \DRLM using \num{1258} queries from four benchmark datasets spanning diverse domains: \texttt{MMLU}~\cite{hendrycks2020measuring} (biology, world history, geography), \texttt{GSM8K}~\cite{cobbe2021gsm8k}, \texttt{CommonsenseQA}~\cite{talmor2018commonsenseqa}, and \texttt{TruthfulQA}~\cite{truthful}. These queries are grouped into \num{6} semantic categories used for class-conditioned modeling. Categories capture semantic query types (not dataset identities) and are assigned via embedding-based classification, mitigating dataset-specific bias. To ensure balanced workloads, \texttt{GSM8K}, \texttt{CommonsenseQA}, and \texttt{TruthfulQA} are subsampled to \num{200} queries each. All datasets provide ground-truth answers, enabling evaluation of response quality across heterogeneous query types. 
% Table~\ref{tab:dataset} summarizes the workload composition. 
Query arrivals follow a Poisson process with rate \mbox{$\lambda \in\{0.5, 1, 2\}$}, modeling varying workload intensities.
\begin{figure*}[!t]
    \centering
    \subfloat[Accuracy of LLMs across queries.\label{fig:acc_models}]{
        \includegraphics[width=0.5\linewidth]{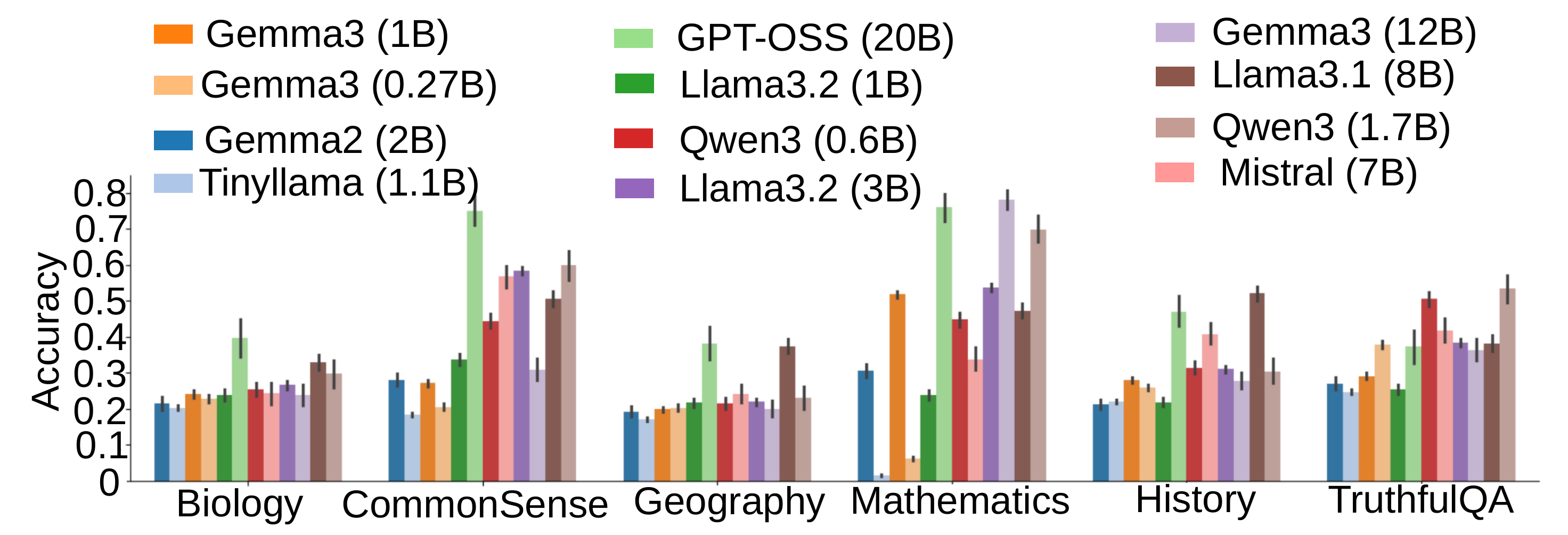}
    }\hfill
    \subfloat[RF query classification.\label{fig:ds_pred}]{
        \includegraphics[width=0.22\linewidth]{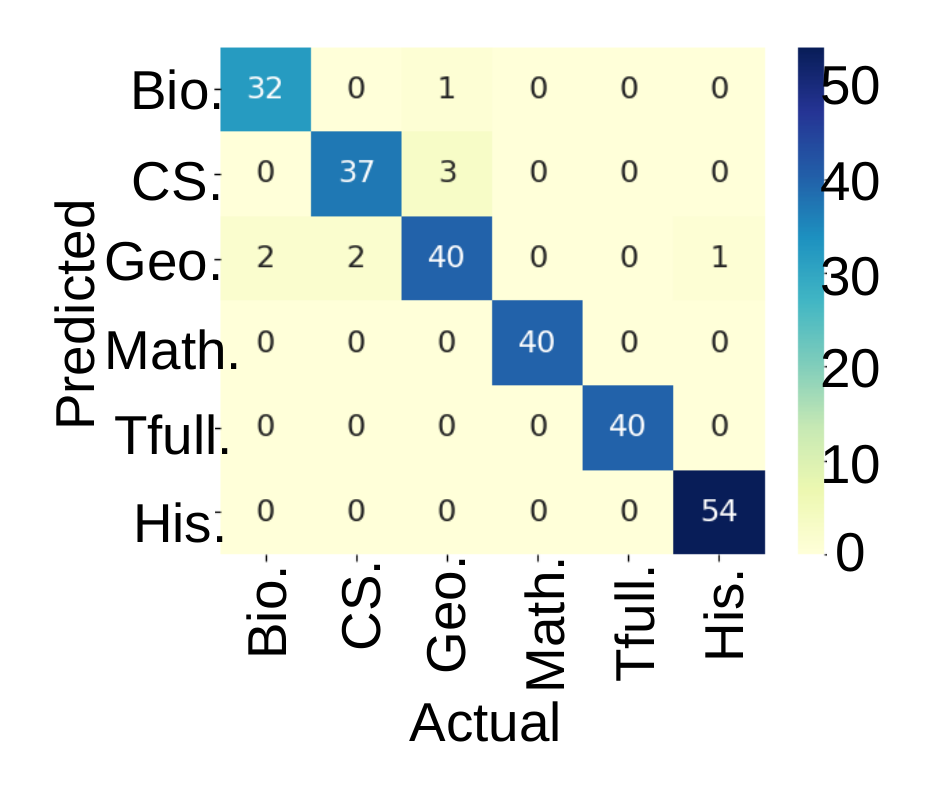}
    }\hfill
    \subfloat[LGBM inference time prediction.\label{fig:inf_pred}]{
        \includegraphics[width=0.22\linewidth]{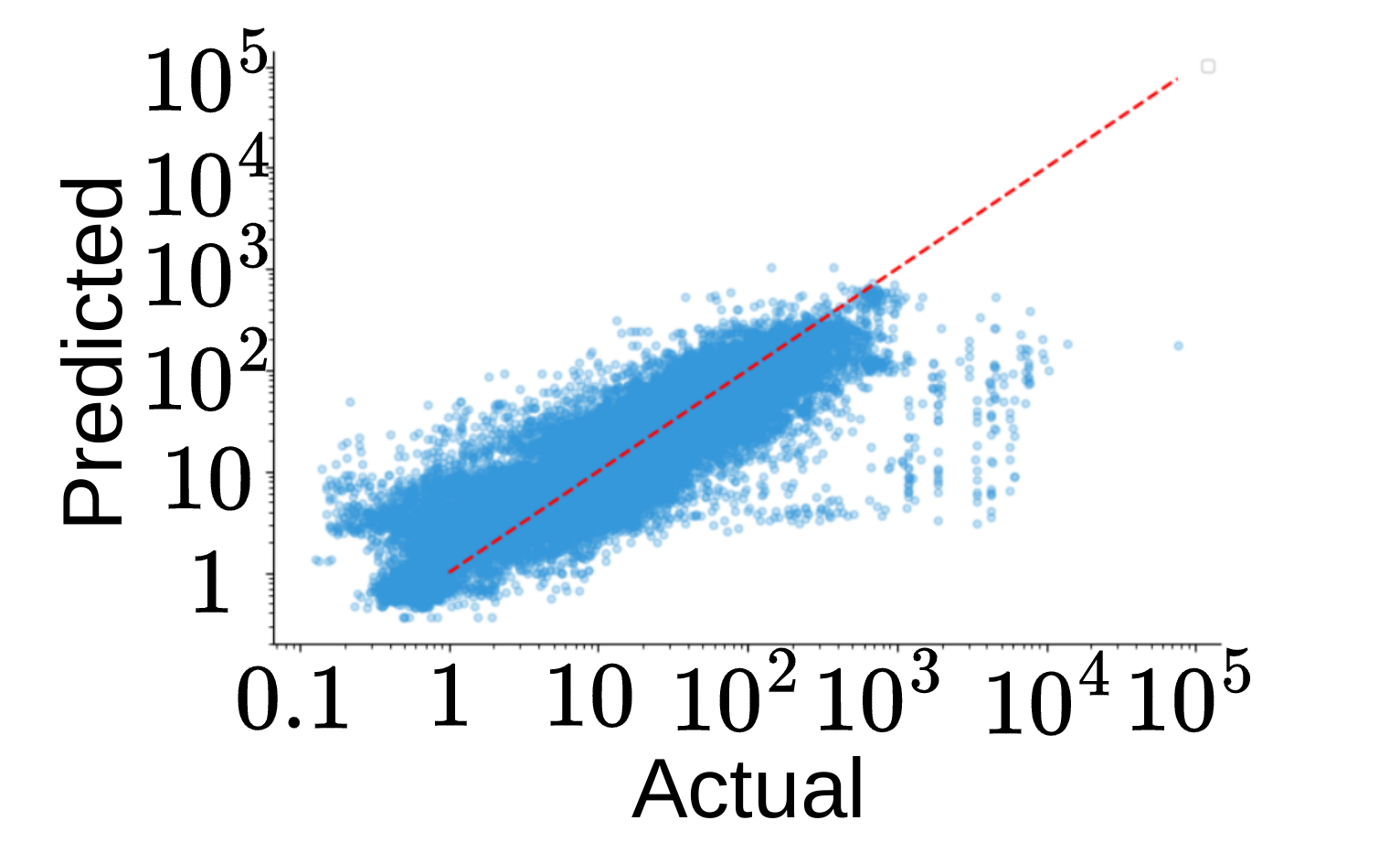}
    }
    \vspace{-5pt}
    \caption{\small{Predictive modeling performance in \DRLM.}}
    \vspace{-.6cm}
    \label{fig:pred_models}
\end{figure*}
\subsubsection{Benchmarking dataset}
We construct a benchmarking dataset with \num{223835} measurements spanning queries, model configurations, quantization levels, and edge devices. Each record corresponds to a query-model-device tuple and includes inference latency, resource utilization (CPU/GPU/memory), and response quality, enabling supervised training of the classifier and predictors. Fig.~\ref{fig:benchmark} summarizes key distributions.
Kernel Density Estimation (KDE) plots (Figs.~\ref{dataset-complexity}-\ref{dataset-token}) reveal multi-modal variability in query complexity and length. Fig.~\ref{dataset-accuracy} shows a bimodal accuracy distribution due to binary and soft-scored tasks, while Fig.~\ref{dataset-inf} exhibits heavy-tailed latency (log scale), reflecting heterogeneity across models and devices. Quality labels are binarized using a \num{0.5} threshold to ensure consistent supervision for class-conditioned estimation.
%Overall, the dataset captures diverse query characteristics and latency-quality trade-offs, forming the empirical basis for learning query-conditioned orchestration policies.
%%
\subsubsection{Query classifier}
We encode each query via \texttt{all-MiniLM-L6-v2} sentence encoder, a \num{6}-layer Transformer producing \num{384}-dimensional embeddings. A Random Forest classifier (\num{200} trees, max depth \num{20}) maps embeddings to \num{6} semantic classes, enabling low-dimensional abstraction of query semantics. For low-confidence or out-of-distribution queries, the system falls back to a predefined class associated with high-capability \texttt{GPT-OSS 20B}, which performs well across query categories (Fig.~\ref{fig:acc_models}), ensuring reliable behavior under uncertainty.
%%%
\subsubsection{Quality estimator} 
We estimate response quality using class-conditioned empirical statistics derived from the benchmarking dataset. Specifically, model performance is aggregated per query class, providing stable estimates of expected accuracy. The estimator is implemented as a precomputed class-model lookup table and queried at runtime, enabling constant-time evaluation of candidate assignments.
%%%
\subsubsection{Latency predictor} We model inference latency using a gradient-boosted regressor (LightGBM) trained on query, model, and device features. The model uses \num{600} trees, \num{63} leaves, and a learning rate of \num{0.05}. Input features include: \emph{(i)} query structure (length, token/word/sentence counts, lexical diversity), \emph{(ii)} model configuration (family, parameter scale, quantization, layers, attention heads), and \emph{(iii)} edge device characteristics (hardware type, CPU/GPU execution). 
%%%
\subsubsection{DRL training} We train the orchestration policy using PPO on trajectories generated by the predictors, avoiding costly LLM execution while preserving system dynamics. The query set is split into disjoint training (\qty{80}{\percent}) and evaluation (\qty{20}{\percent}) subsets. The actor-critic network shares a two-layer MLP encoder (\num{128} units per layer). The actor is factorized into three heads, and the critic outputs a scalar value estimate. Training runs for \num{1000} episodes, each simulating \num{100} requests with exponential inter-arrivals (rate \num{1}). We use Adam (lr $3\times10^{-4}$), discount factor \num{0.95}, entropy coefficient \num{0.02}, clipping \num{0.2}, and \num{6} update epochs per rollout. Reward weights are set to \mbox{$\alpha=0.8$} and \mbox{$\beta=0.3$}, empirically calibrated to balance quality and latency under normalized rewards.
% For comparison, the DQN baseline uses a three-layer MLP (\num{128} units each, ReLU) trained for \num{1000} episodes with minibatches of \num{64} from a replay buffer of size \num{10000}. Exploration follows an $\epsilon$-greedy policy decaying from \num{1.0} to \num{0.05} (decay \num{0.995}). Optimization uses Adam (lr $10^{-3}$) with discount factor \num{0.95}, and a target network updated every \num{10} episodes.
%%%%
\subsubsection{Comparison methods}
We compare \DRLM against three baselines and two state-of-the-art methods under the same testbed, deployment, and workload:
\begin{enumerate*}[label=(\emph{\roman*})]
\item \emph{Random:} uniformly samples a feasible $(m,e)$, ignoring query and system state.
\item \emph{High-Acc:} selects $\arg\max_m \mathcal{A}(c,m)$, ignoring latency and load.
\item \emph{Fastest:} selects $(m,e)$ minimizing predicted inference latency, ignoring quality.
\item \emph{RouteLLM-style:} adapted~\cite{ong2025routellm} by defining a single strong model (\texttt{GPT-OSS} \qty{20}{B}) and a weak pool (all others). Queries are routed to the strong model if its predicted advantage exceeds \num{0.5}, otherwise to the best weak model based on $\mathcal{A}(c,m)$.
\item \emph{OptLLM-style:} adapted~\cite{liu2024optllm} by replacing cost with latency and restricting decisions to deployed $(m,e)$ pairs; selection is performed from the Pareto frontier of quality-latency trade-offs.
% \item \emph{DQN:} replaces PPO with a Deep Q-Network (DQN)-based agent under the same state and reward formulation.
\end{enumerate*}
%%%%
\subsubsection{Evaluation metrics}
We evaluate all methods using the following metrics:
\begin{enumerate*}[label=(\emph{\roman*})]
\item \emph{Response quality:} average correctness against ground truth, capturing task-level accuracy across heterogeneous queries.
\item \emph{Inference latency:} model execution time, reflecting efficiency of selected model-device configurations.
\item \emph{Waiting time:} queueing delay before execution, capturing load-induced congestion.
\item \emph{Orchestration overhead:} per-query decision time of the orchestration policy.
\end{enumerate*}
% \za{Each policy is evaluated over 10 runs with seeds 1000–1009. Results are averaged across runs and reported with standard deviation.}

\section{Experimental Results}
\label{sec:EvaluationSetup}
%%%%%%%%%%%%%%%%%%%%%%%%%%%%%%%%%
%%
\begin{figure}[!t]
    \centering
     \subfloat[Latency predictor feature importance.]{
        \includegraphics[width=0.5\linewidth]{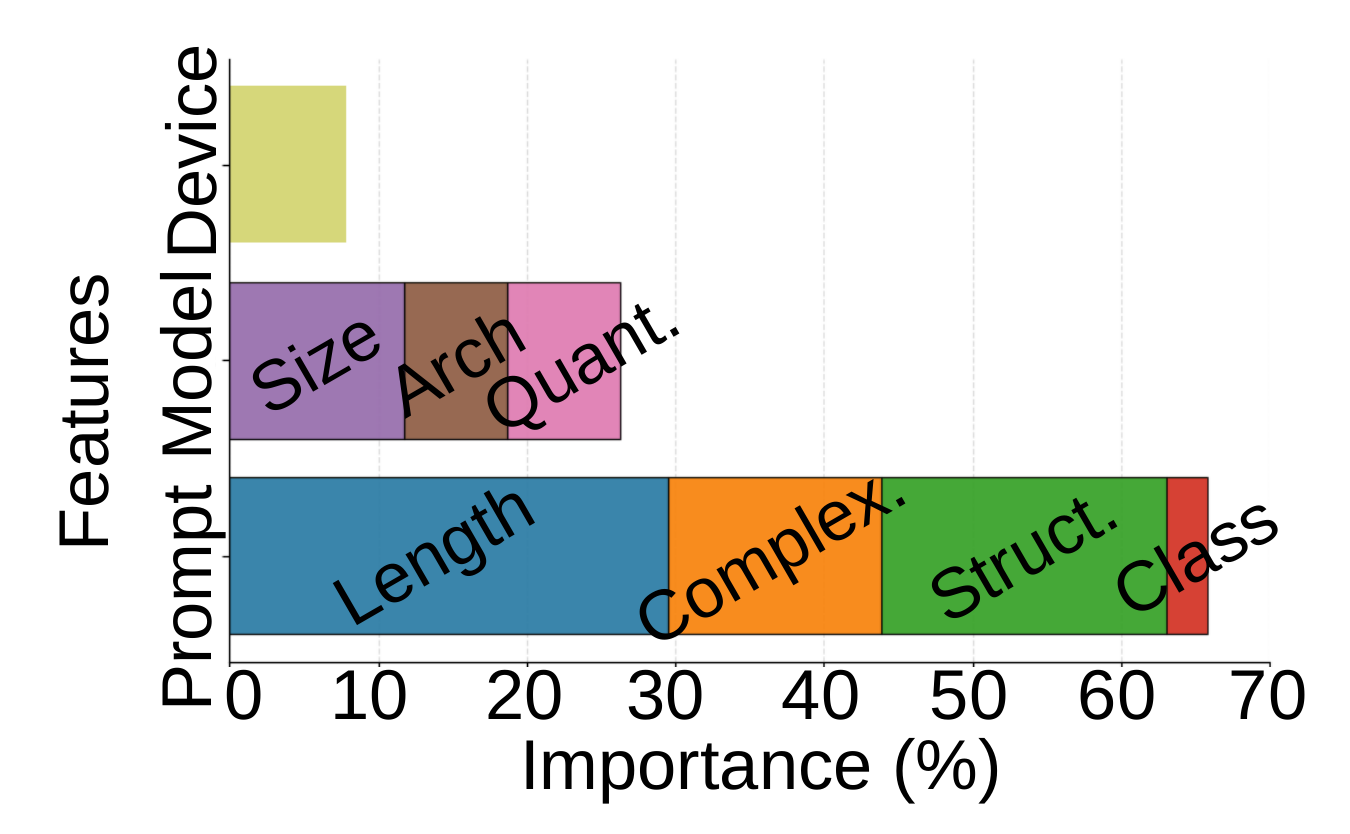}
        \label{fig:feat_imp}
    }\hfill
    \subfloat[Reward and loss convergence.]{
        \includegraphics[width=0.42\linewidth]{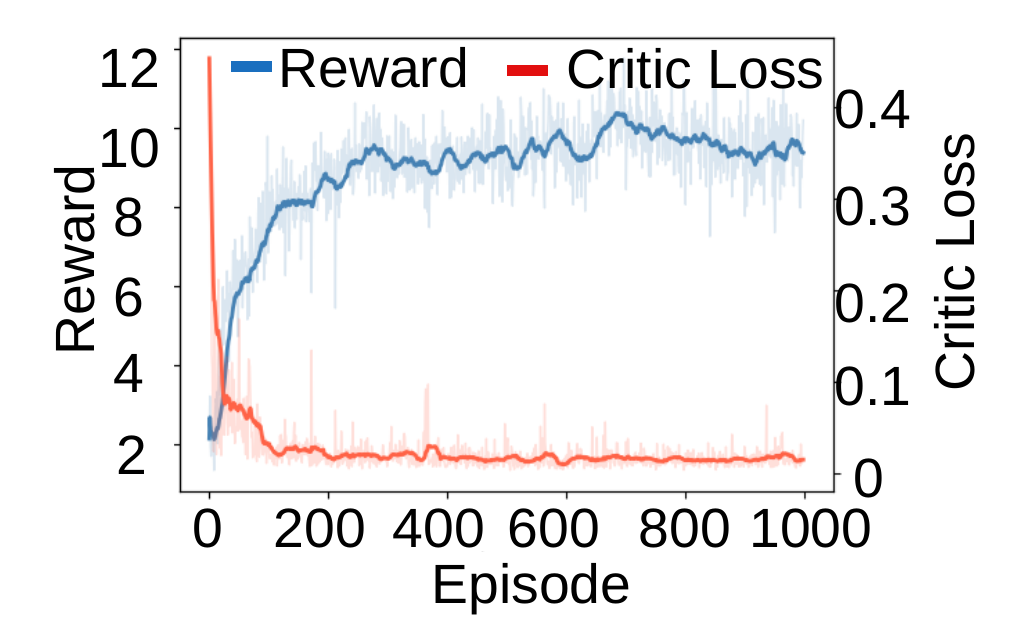}
        \label{fig:reward}
    }\hfill
    \vspace{-5pt}
    \caption{\small{Training dynamics in \DRLM.}}
    \label{fig:orch-ppo}
    \vspace{-25pt}
\end{figure}
\begin{figure}[t!]
    \centering
        \subfloat[Orchestration time.]{
        \includegraphics[width=0.5\linewidth]{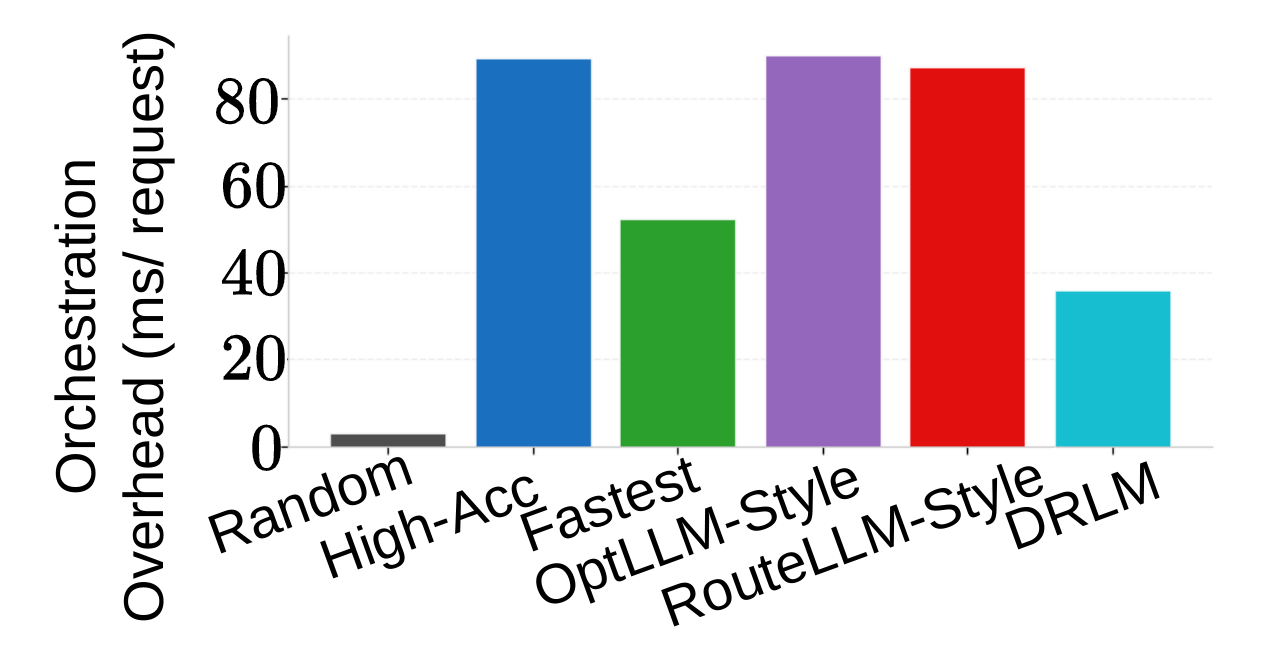}
        \label{fig:overhead}
    }
    \hfill
    \subfloat[Waiting time (log-scale).\label{fig:sota_cdf}]{
        \includegraphics[width=0.45\linewidth]{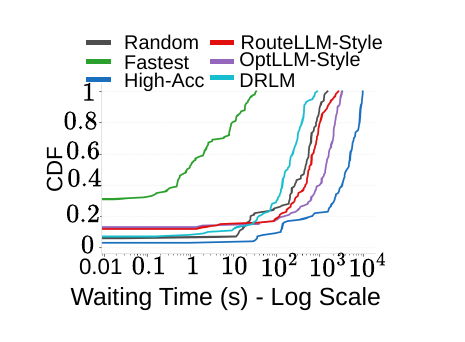}}
    \hfill
    \vspace{-5pt}
    \caption{\small{Orchestration and waiting time across baselines.}}
    \label{fig:sota-comp1}
\end{figure}
%%%%%%%%%%%%
%%
\begin{figure*}[!t]
    \centering
    \subfloat[Model-device selection distribution.\label{fig:sota_dist}]{
        \includegraphics[width=0.3\linewidth]{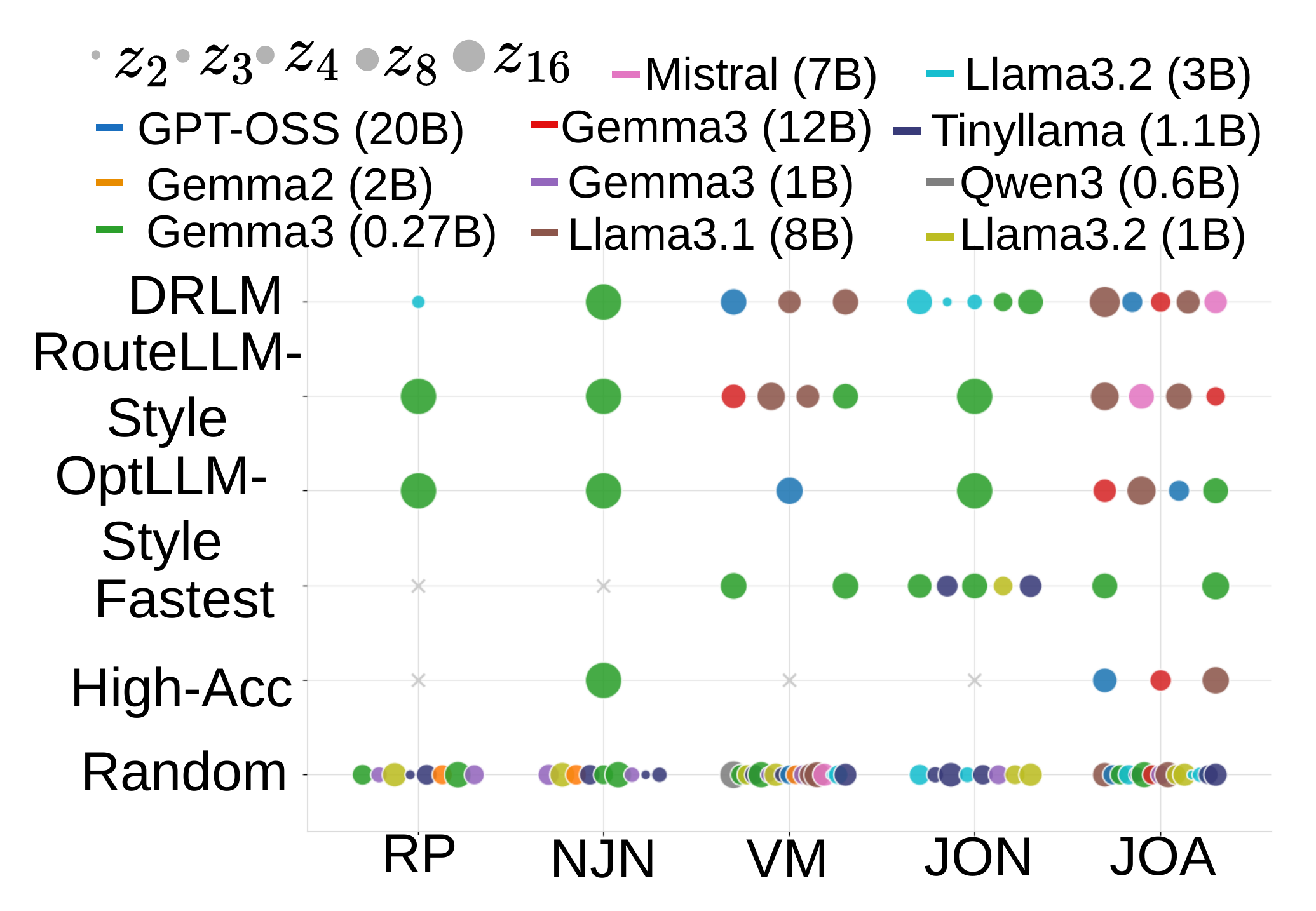}}
    \hfill
    \subfloat[Average response quality.\label{fig:sota_acc}]{
        \includegraphics[width=0.32\linewidth]{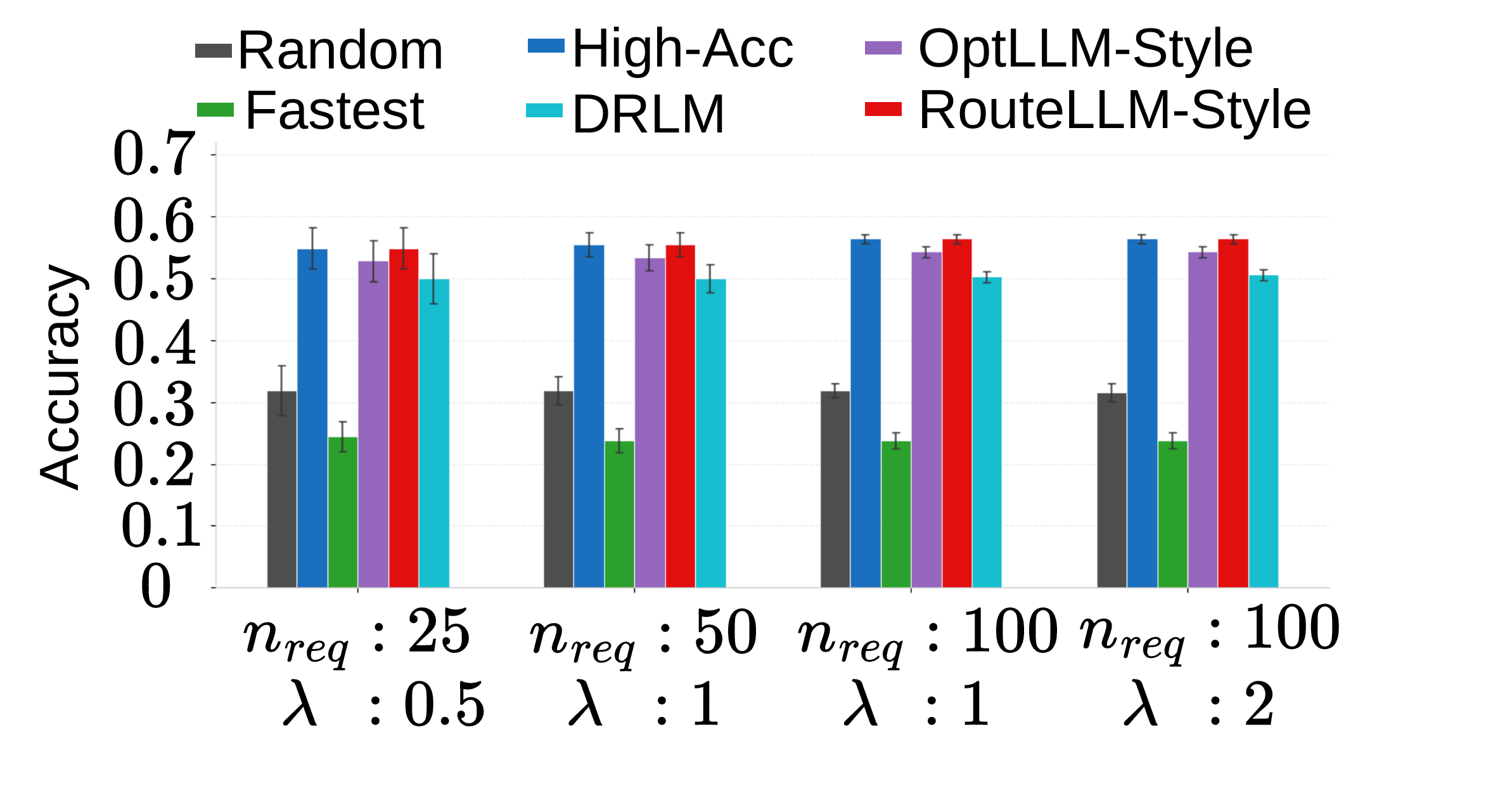}}
    \hfill
    \subfloat[End-to-end latency (log-scale).\label{fig:sota_inf}]{
        \includegraphics[width=0.32\linewidth]{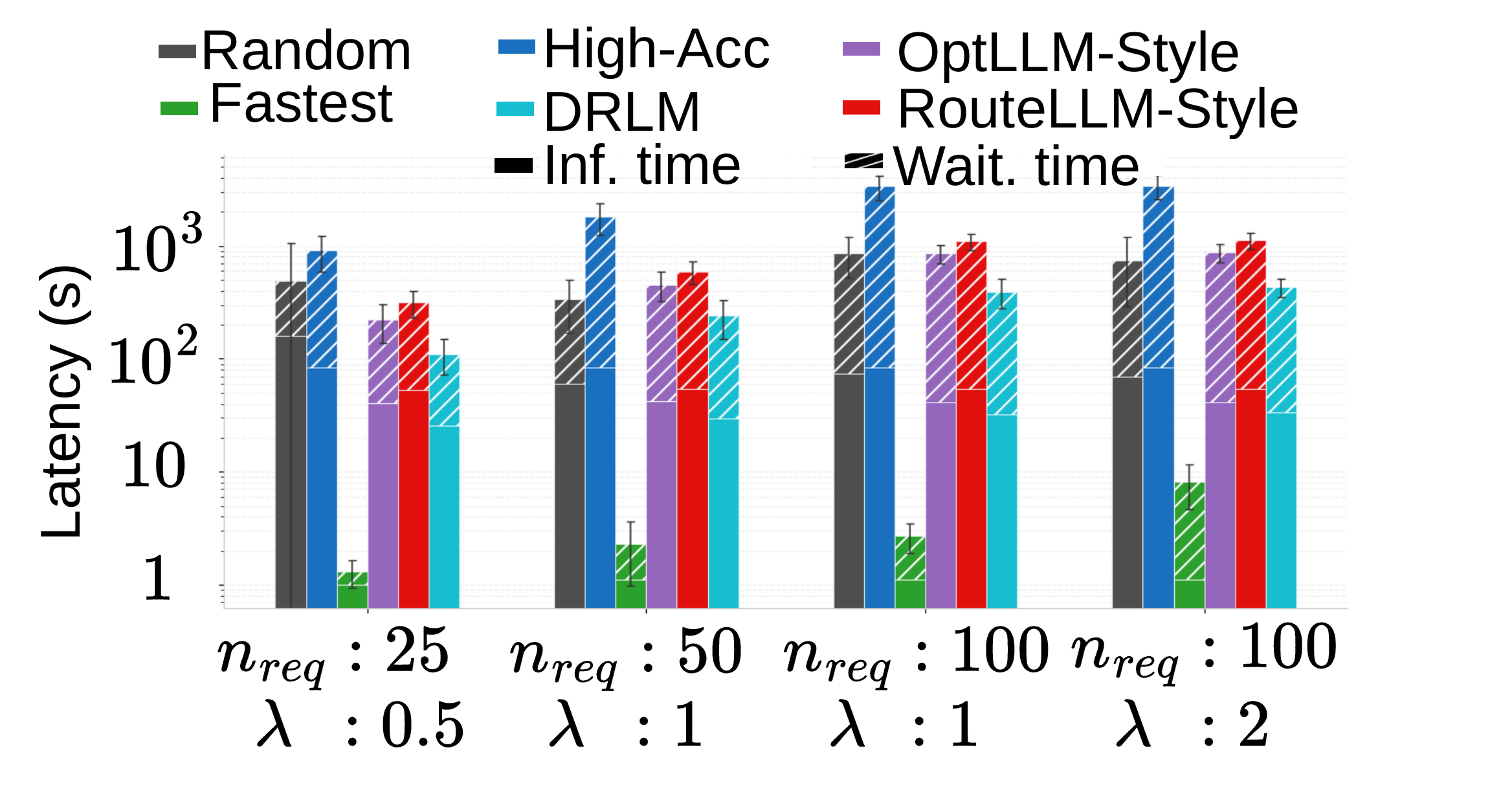}}
        \vspace{-5pt}
    \caption{\small{Comparison of model selection, response quality, and end-to-end latency across baselines under varying workload intensities.}}
    \label{fig:sota-comp2}
    \vspace{-.4cm}
\end{figure*}
This section evaluates \DRLM against three baselines and two state-of-the-art methods on the same edge testbed, reporting the average and standard deviation over \num{10} independent runs.
\subsubsection{Predictive modeling analysis}
Fig.~\ref{fig:acc_models} shows variability in model accuracy across query categories, with no single model dominating across all tasks. This non-uniform behavior confirms that model effectiveness is inherently query-dependent, motivating class-conditioned quality estimation rather than global model ranking. The RF-based query classifier (Fig.~\ref{fig:ds_pred}) achieves high consistency in mapping queries to semantic categories, as reflected by the strongly diagonal confusion matrix with limited cross-class leakage. This behavior stems from the semantic coherence of queries within each category, allowing embedding-based representations to effectively cluster similar queries. Rather than enforcing strict boundaries, the classifier provides a stable abstraction of query types suitable for downstream class-conditioned estimation. 
%%%%%%%%%%%%
The latency predictor (Fig.~\ref{fig:inf_pred}) achieves consistent alignment between predicted and measured inference times on a log scale (\mbox{$R^2=0.75$}), capturing relative latency differences across model-device configurations. This level of accuracy is sufficient for query orchestration, as the policy depends on preserving the ordering of candidate configurations rather than minimizing absolute prediction error. Its inference overhead is negligible (\qty{0.78}{ms} per instance), enabling real-time evaluation of candidate assignments. Feature importance analysis (Fig.~\ref{fig:feat_imp}) shows that prompt-level features dominate latency prediction, with input length and structural complexity contributing the largest share, followed by model configuration, while device features have a lower impact. This is because devices are partially entangled with model deployment, leading to shared explanatory power; thus, model configuration captures most of the variance associated with execution cost.
%%%
\subsubsection{DRL convergence and orchestration time}
Fig.~\ref{fig:reward} shows the training dynamics of the PPO-based \DRLM agent. The reward increases rapidly within the first \num{200} episodes and stabilizes at a high-value plateau, indicating fast and stable convergence. In parallel, the critic loss drops sharply and remains near zero, confirming accurate value estimation and low training variance. Fig.~\ref{fig:overhead} compares orchestration overhead across methods. \DRLM achieves approximately \qty{35}{ms/query}, reducing decision latency by about \qty{60}{\percent} compared to OptLLM-style and RouteLLM-style, and by about \qty{30}{\percent} compared to Fastest. High-Acc incurs similar overhead to optimization-based methods due to full candidate evaluation.
This efficiency is enabled by the factorized policy design, which decomposes decision-making into sub-actions, reducing the effective action space and avoiding exhaustive evaluation.
%%%
\subsubsection{Comparison with baselines}
Fig.~\ref{fig:sota_dist} shows the distribution of query assignments across model-device configurations. \DRLM learns a load-aware policy that distributes queries across both lightweight and high-capacity configurations, assigning simple queries to quantized models on constrained devices (RP, NJN) while reserving larger models on powerful nodes (VM, JON, JOA) for complex queries. In contrast, High-Acc concentrates assignments on large models, creating hotspots, while Fastest collapses to aggressively quantized configurations. Random exhibits no structure. OptLLM- and RouteLLM-style partially balance this trade-off but remain less adaptive to device-level heterogeneity. 
Fig.~\ref{fig:sota_cdf} shows the CDF of waiting time (log scale) in different schemes. \DRLM shows a left-shift with a steeper rise, indicating lower median and tail latency. In contrast, High-Acc, OptLLM, and RouteLLM show heavy-tailed behavior, with a large fraction of queries experiencing delays above \qty{100}{\second}. 
Fastest achieves low waiting time but at the cost of degraded quality (Fig.~\ref{fig:sota_acc}).

Fig.~\ref{fig:sota_acc} reports average response quality under increasing workload (\mbox{$n_{\text{req}}\in\{25,50,100\}$}, \mbox{$\lambda\in\{0.5,1,2\}$}). High-Acc and RouteLLM achieve the highest accuracy (\numrange{0.54}{0.56}), followed by OptLLM. \DRLM maintains stable performance around \numrange{0.50}{0.52}, outperforming Fastest and Random. 
%These accuracy levels are expected given the difficulty of the benchmarks and the inclusion of lightweight quantized models; as shown in Fig.~\ref{fig:acc_models}, no model uniformly achieves high accuracy across query categories. 
\DRLM exhibits minor variation across workload intensities, indicating stable behavior under increasing load. Fig.~\ref{fig:sota_inf} shows end-to-end latency (log scale) under varying workloads. As load increases, latency rises sharply for accuracy-oriented baselines due to queuing effects caused by over-selection of large models. In contrast, \DRLM mitigates congestion via state-aware orchestration, distributing queries across model-device configurations. Under the highest load (\num{100} requests, \mbox{$\lambda=2$}), \DRLM reduces inference time by \qty{19.2}{\percent} and \qty{38}{\percent} compared to OptLLM and RouteLLM, and reduces waiting time by \qty{52}{\percent} and \qty{62}{\percent}, with at most \qty{8}{\percent} accuracy loss.

%%%%%%%%%%%%%%%%%%%%%%%%%%Extra Result plan% %%%%%%%%%%%%%%%%%%%%%%%%%%%
% %scenario1: result accuracy and total time when increasing the number of requests (fixed rate)
% %scenario2: result accuracy and time when increasing rate 
% %predictors (1- knn and benchmarks and inference time)
% % Shap result for inference time
% % distributions (model, device heatmaps) when scaling requests and bar charts for showing qeueing distributions
% % CDF of waiting time
% % how many of request serveed without q
% % Execution time and batching collection per policy 
% % Resource utilization
% % Query distribution on models in different approaches
% % routing decision distribution
% Ablation: without classifier / without factorization.
% discussion of bias/coverage. for dataset
%not fixed deployment (add scaling/scheduling decisions)

\section{Conclusion and Future Work}\label{sec:Conclusion} 
This paper presented \DRLM, a DRL-based framework for fine-grained LLM query orchestration in heterogeneous edge clusters. \DRLM jointly models \emph{(i)} query semantics via embedding-based classification, \emph{(ii)} class-conditioned quality and feature-driven latency prediction, and \emph{(iii)} dynamic system state (resource utilization and queuing), enabling state-aware model-device selection. We constructed a large-scale benchmarking dataset with \num{223835} measurements, capturing query-dependent latency-quality trade-offs across models and edge devices. Evaluation on a \num{64}-node edge testbed shows that \DRLM shortens inference latency by up to \qty{51}{\percent} and queuing delay by up to \qty{67}{\percent}. Future work will extend \DRLM toward multi-objective optimization, incorporating cost and energy.
% \section*{Acknowledgment}
% The financial support of the Austrian Federal Ministry for Digital and Economic Affairs, the National Foundation for Research, Technology and Development, the Christian Doppler Research Association is gratefully acknowledged. Christian Doppler Laboratory ATHENA: \url{https://athena.itec.aau.at/}.
\balance
\bibliographystyle{./bibliography/IEEEtran}
\bibliography{./bibliography/IEEEabrv}
\end{document}